\documentclass[sts]{imsart}

\RequirePackage{amsthm,amsmath,amsfonts,amssymb}
\RequirePackage[numbers,sort&compress]{natbib}
\RequirePackage[colorlinks,citecolor=blue,urlcolor=blue]{hyperref}
\RequirePackage{graphicx}
\usepackage{comment}
\usepackage{bm}

\newcommand{\blA}{\bm{A}}

\newcommand{\blD}{\bm{D}}
\newcommand{\blE}{\bm{E}}

\newcommand{\blM}{\bm{M}}

\newcommand{\blQ}{\bm{Q}}

\newcommand{\blS}{\bm{S}}

\newcommand{\blU}{\bm{U}}

\newcommand{\blW}{\bm{W}}
\newcommand{\blX}{\bm{X}}

\newcommand{\blh}{\bm{h}}

\newcommand{\blx}{\bm{x}}

\newcommand{\bggamma}{\bm{\gamma}}

\newcommand{\AAA}{\mbox{$\mathbf A$}}

\newcommand{\ZZ}{\mbox{$\mathbf Z$}}

\newcommand{\DD}{\mbox{$\mathbf D$}}

\newcommand{\so}{\mbox{$\widehat{\mathcal V}_{o, cons}$}}
\newcommand{\sone}{\mbox{$\widehat{\mathcal V}_{grp, anch}$}}
\newcommand{\stwo}{\mbox{$\widehat{\mathcal V}_{grp, solo}$}}
\newcommand{\sthree}{\mbox{$\widehat{\mathcal V}_{out}$}}

\newcommand{\sot}{\mbox{$\mathcal V_{o, cons}$}}
\newcommand{\sonet}{\mbox{${\mathcal V}_{grp, anch}$}}
\newcommand{\stwot}{\mbox{${\mathcal V}_{grp, solo}$}}
\newcommand{\sthreet}{\mbox{$\mathcal V_{out}$}}

\newcommand{\bbb}{\mbox{$\boldsymbol{\beta} $}}

\newcommand{\ddd}{\mbox{$\boldsymbol{\delta} $}}
\newcommand{\zzz}{\mbox{$\boldsymbol{\zeta} $}}
\newcommand{\bbbb}{\mbox{$\boldsymbol{\beta} $}}
\newcommand{\BB}{\mbox{$\mathbf{B} $}}

\newcommand{\DDDD}{\mbox{$\boldsymbol{\Delta} $}}

\newcommand{\Orth}{\mbox{$\mathcal{O}_r $}}

\newcommand{\bbR}{\mbox{$\mathbb{R}$}}
\newcommand{\trans}{^{\top}}

\newcommand{\cV}{\mbox{$\mathcal V$}}

\newcommand{\SSS}{\mbox{$\mathbf S$}}

\newcommand{\embSVD}{\tilde{\blX}^{}}
\newcommand{\embSVDa}{\tilde{\blX}^{\textrm{a}}}
\newcommand{\embsvda}{\tilde{\blx}^{\textrm{a}}}
\newcommand{\embsvd}{\tilde{\blx}^{}}
\newcommand{\embsvdaa}{\tilde{\blx}^{\textrm{a2}}}

\newcommand{\cP}{\mathcal{P}}

\newcommand{\gginithat}{\hat{\boldsymbol{\gamma}}_g^{\textrm{init}}}
\newcommand{\ginithat}{\hat{\boldsymbol{\gamma}}^{\textrm{init}}}

\newcommand{\subso}{\scriptscriptstyle{\widehat{\mathcal{V}}_{o, cons}}}
\newcommand{\subsot}{\scriptscriptstyle{\mathcal{V}_{o, cons}}}

\newcommand{\subtrans}{\scriptscriptstyle{\widehat{\mathcal{V}}_{trans}}}

\newtheorem{remark}{Remark}
\newtheorem{assumption}{Assumption}
\newcommand{\pr}{\mbox{$\mathbb P$}}

\startlocaldefs
\theoremstyle{plain}

\newtheorem{theorem}{Theorem}[section]
\newtheorem{lemma}[theorem]{Lemma}
\theoremstyle{definition}

\endlocaldefs

\begin{document}

\begin{frontmatter}
\title{Transfer Learning with Network Embeddings under Structured Missingness}

\begin{aug}
\author[A]{\fnms{Mengyan}~\snm{Li}\thanksref{t1}\ead[label=e1]{mengyanli@bentley.edu}},
\author[B]{\fnms{Xiaoou}~\snm{Li}\thanksref{t1}\ead[label=e2]{lixx1766@umn.edu }},
\author[C]{\fnms{Kenneth D}~\snm{Mandl}\ead[label=e3]{Kenneth.Mandl@childrens.harvard.edu}},
\and
\author[D]{\fnms{Tianxi}~\snm{Cai}\ead[label=e4]{tcai@hsph.harvard.edu} \thanks{\textbf{Tianxi Cai is the Corresponding author }}}

\thankstext{t1}{Mengyan Li and Xiaoou Li are co-first authors and contributed equally to this work.}

\address[A]{Mengyan Li is Assistant Professor, Department of Mathematical Sciences,
Bentley University, Waltham, USA\printead[presep={\ }]{e1}.}

\address[B]{Xiaoou Li is Associate Professor, School of Statistics,
University of Minnesota, Minneapolis, USA\printead[presep={\ }]{e2}.}

\address[D]{Kenneth D. Mandl is Professor, Chair in Biomedical Informatics and Population Health, Harvard Medical School, Boston, USA\printead[presep={\ }]{e3}.}

\address[D]{Tianxi Cai is John Rock Professor, Harvard T.H. Chan School of Public Health,
Harvard University, Boston, USA\printead[presep={\ }]{e4}.}

\end{aug}

\begin{abstract}
Modern data-driven applications increasingly rely on large, heterogeneous datasets collected across multiple sites. Differences in data availability, feature representation, and underlying populations often induce structured missingness, complicating efforts to transfer information from data-rich settings to those with limited data. Many transfer learning methods overlook this structure, limiting their ability to capture meaningful relationships across sites.
We propose TransNEST ({\bf Trans}fer learning with {\bf N}etwork {\bf E}mbeddings under {\bf ST}ructured missingness), a framework that integrates graphical data from source and target sites with prior group structure to construct and refine network embeddings. TransNEST accommodates site-specific features, captures within-group heterogeneity and between-site differences adaptively, and improves embedding estimation under partial feature overlap. We establish the convergence rate for the TransNEST estimator and demonstrate strong finite-sample performance in simulations.
We apply TransNEST to a multi-site electronic health record study, transferring feature embeddings from a general hospital system to a pediatric hospital system. Using a hierarchical ontology structure, TransNEST improves pediatric embeddings and supports more accurate pediatric knowledge extraction, achieving the best accuracy for identifying pediatric-specific relational feature pairs compared with benchmark methods.

\end{abstract}


\end{frontmatter}

\section{Introduction}
\subsection{Background}
Modern data-driven applications increasingly involve learning from multiple data sources or sites, where feature sets only partially overlap, and data availability varies across sources. Such settings arise in many domains, including multilingual representation learning, cross-platform recommendation systems, and multi-site clinical studies where measurement protocols and recorded covariates differ across centers.
A fundamental challenge in these problems is structured missingness induced by heterogeneous feature spaces and population-level differences across sites, which complicates transfer learning from information-rich sources to data-limited targets.

In many multi-site settings, a subset of features admits known one-to-one correspondences across sites, while the remaining features are site-specific but organized through a shared group or hierarchical structure. For example, in multilingual representation learning, sites may correspond to corpora in different languages with partially overlapping vocabularies or known cross-lingual alignments. In cross-platform recommendation systems, sites may correspond to different platforms with overlapping item sets or externally matched items. In both cases, features are connected through network-structured representations derived from co-occurrence or similarity information, and group structure encodes shared semantic or taxonomic relationships.

Another example is multi-site knowledge extraction from electronic health record (EHR) data.
Advances in EHR systems have enabled the use of large-scale real-world clinical data for data-driven biomedical research, supporting tasks such as clinical knowledge extraction, risk prediction, and treatment response assessment \citep[e.g.][]{liu2013information, chen2019robustly, goldstein2016opportunities, sheu2023ai}. Despite this promise, the complex structure of EHR data poses substantial modeling challenges. EHR features are high-dimensional and relational, comprising diagnoses, medications, laboratory tests, and procedures. Representation learning has therefore emerged as an effective approach for modeling EHR data, enabling low-dimensional embeddings that capture feature relationships and support downstream tasks \citep{si2021deep}.
Beyond high dimensionality, EHR data exhibit pronounced heterogeneity across patient subpopulations and healthcare systems. In particular, representations learned from general or adult populations may transfer poorly to pediatric settings due to biological, pharmacological, and clinical differences, as well as pediatric-specific vocabularies \citep{fernandez2011factors}. Additionally, approximately half of pediatric healthcare visits are preventive well-checks \citep{gracy2012content, joseph2015clinical}, contributing to the unique sparsity of pediatric EHR data. Although shared hierarchical ontology structures can potentially increase transferability between adult and pediatric populations, these differences induce structured missingness and site-specific feature behavior, limiting the effectiveness of standard transfer learning methods that assume shared feature spaces.

These challenges motivate transfer learning frameworks that can exploit network structure, handle partial feature alignment, and incorporate shared group information across heterogeneous multi-site data. To this end, we propose a {\bf Trans}fer learning with {\bf N}etwork {\bf E}mbeddings under {\bf ST}ructured missingness (TransNEST) method, by combining site-specific network embeddings with prior group structure to guide selective information sharing. TransNEST accommodates partially aligned feature spaces by allowing site-specific features, while adaptively modeling heterogeneity both within groups and across sites. As a result, it enables more efficient and accurate knowledge transfer to sparse, heterogeneous target populations.

\subsection{Related Literature and Our Contributions}
Learning from multi-site data with heterogeneous feature spaces and structured missingness has been studied across several methodological lines, including matrix completion, multi-task learning, representation learning, and transfer learning. While these approaches offer partial solutions, they typically fall short in settings characterized by partial feature alignment, multi-block missingness patterns, and complex cross-site heterogeneity, which arise in many real-world applications.

Classical matrix completion methods \citep[e.g.][]{candes2010power, cai2016matrix, mazumder2010spectral, rohde2011estimation}
typically assume entries are missing completely at random, an assumption violated in multi-site settings
where missingness is driven by systematic differences in feature availability across sites.
Structured matrix completion relaxes this assumption for structured missingness, where an entire submatrix block is unobserved \citep{cai2016structured}, but it does not directly accommodate more general multi-block missingness patterns common in multi-source data.
More recent work considers matrices with overlapping sub-blocks \citep{zhou2023multi},
yet enforces identical embeddings for overlapping features, which can be overly restrictive
when feature semantics or relationships differ across sites.

Related challenges have also been studied from a multi-task learning perspective.
For example, \cite{sui2025multi} propose a two-step framework for heterogeneous multi-source
block-wise missing data, addressing task-level heterogeneity but operating under a different
modeling paradigm than the network-embedding–based transfer framework developed here.
More broadly, multi-task with shared-representation methods, including multi-task PCA and related
approaches \citep[e.g.][]{fan2019distributed, yamane2016multitask, li2024knowledge},
aim to borrow strength across tasks by learning shared latent structure.
However, these methods generally assume a common feature space and do not accommodate
site-specific features or structured missingness induced by heterogeneous data collection processes.

Several lines of statistical work incorporate external structural information to enhance embedding learning or facilitate knowledge transfer. \cite{binkiewicz2017covariate} proposes covariate-assisted spectral clustering that jointly leverages graph structure and node covariates to recover latent communities. \cite{bi2017group} incorporates group structure among rows and columns, enabling matrix completion under missing not at random mechanisms. However, these two frameworks are developed for single network settings and do not consider block-wise missingness induced by heterogeneous feature spaces across multiple sites. \cite{shi2021spherical} develops a spherical regression framework to learn a mapping between two embedding spaces under mismatches and can incorporate group information, but it imposes strong correspondence constraints and does not address multi-site transfer with multi-block missingness.

Complementing these statistical approaches, a large body of work studies knowledge transfer through learned representations in deep models.
Deep representation learning and transfer learning methods\citep[e.g.][]{hinton2006reducing, hinton2006fast, bengio2012deep} learn transferable features from unlabeled data and adapt source-trained representations to target settings. While related in spirit, these approaches typically do not explicitly model structured missingness, and often lack theoretical guarantees. More specialized deep transfer methods, such as distribution alignment \citep{zhuang2015supervised} or similarity-induced embeddings \citep{passalis2018unsupervised}, alleviate some of these issues but still rely on restrictive alignment assumptions.

TransNEST explicitly accommodates site-specific features, partial feature alignment, and heterogeneous populations by integrating network-structured representations observed at each site with shared group or hierarchical structure. The method adaptively determines which features should borrow strength across sites and/or from the group, and refines embeddings through a procedure that leverages cross-site and within-group homogeneity,  and feature-level priors, such as feature-site weights. This selective information sharing improves embedding estimation efficiency while reducing bias induced by heterogeneous feature behavior.

Importantly, we also develop rigorous theoretical guarantees for TransNEST estimator
under realistic multi-site structures. We establish
error bounds for embedding accuracy and prove that, under mild conditions, TransNEST performs no worse than single-site singular value decomposition (SVD), ensuring robustness against negative transfer.
With intermediate two-to-infinity norm bounds established, these results connect to and extend recent developments in entrywise error bounds for low-rank models
\citep[e.g.][]{abbe2020entrywise, cape2019two, chernozhukov2023inference,
chen2024note}, but accommodate a substantially more complex, heterogeneous, and
multi-block missingness structure. 

In summary, TransNEST provides an adaptive, theoretically grounded solution to transfer learning with network-structured data under structured missingness, with broad applicability across multi-site and multi-source learning problems.

\section{Method} 
\subsection{Notation}
For any positive integer $n$, let $[n]= \{1, \dots, n\}$. For any set $S$, let $|S|$ denote its cardinality. 
For a matrix $\blA = (a_{ij})\in \mathbb{R}^{n\times p}$, 
let $\blA_{i\cdot}$ and $\blA_{\cdot j}$ denote the $i$-th row and the $j$-th column, respectively, and let $\tau_1(\AAA)\geq\cdots\geq \tau_{\min(n,p)}(\AAA)$ denote its singular values. If $\AAA$ is positive semi-definite, then they are also the same as the eigenvalues.
Let $\|\blA\|_F$ and $\|\blA\|_2$ denote the Frobenius and spectral norm of a matrix $\blA$, respectively. In addition, $\|\AAA\|_{2\to\infty}=\max_{i\in[n]}(\sum_{j=1}^p a_{ij}^2)^{1/2}$ denotes the two-to-infity norm of $\blA$, which is the largest row norm of $\blA$. 
For any full rank matrix $\blA$, let $\mathbb{U}(\blA) = \blA(\blA^T\blA)^{-1}$ be its polar factor.
Let $\Orth$ denote the set of all $r\times r$ orthogonal matrices.
We further define the operator $\mathbb{W}(\blX,\tilde{\blX}) = \arg\min_{\blQ\in\Orth}\|\tilde{\blX}-\blX \blQ\|_F = \mathbb{U}(\blX^T\tilde{\blX})$, which is known as the orthogonal procrustes problem in the literature with a closed-form solution \citep{schonemann1966generalized}.
For two sequences $a_n$ and $b_n$, we denote by $a_n=O(b_n)$ if $\lim_{n\rightarrow\infty}|a_n/b_n|<\infty$, $a_n=o(b_n)$ if $\lim_{n\rightarrow\infty}a_n/b_n=0$, and  $a_n=O_p(b_n)$ or $a_n=o_p(b_n)$ if $a_n=O(b_n)$ or  $a_n=o(b_n)$ with a probability approaching $1$. 
We use $a_n \lesssim b_n$ to denote $a_n \leq Cb_n$ for some constant $C>0$,
and use $a_n \sim b_n$ to denote $C\leq a_n/b_n \leq C^{\prime}$ for some constants $C, C^{\prime}>0$.

\subsection{Model Setup}

Let $\blS = (s_{k,i,j})_{k=1,2;\, i,j\in[n]}$ denote the underlying data tensor,
where $k$ indexes the site and $n$ is the total number of distinct features
across sites. For each site $k$, the entry $s_{k,i,j}$ quantifies the
relatedness between features $i$ and $j$. 
Because the sites may not share
identical feature sets, let $\cV_k\subset [n]$ denote the feature indices
observed at each site, with $|\cV_k| = n_k$, $k=1,2$.
The observed site-specific matrix is
$\blS_k = (s_{k,i,j})_{i,j\in\cV_k} \in \bbR^{n_k \times n_k}$,
and for simplicity we assume $\blS_k$ is symmetric.
In practice, $\blS_k$ may be constructed as a Shifted Positive Pointwise Mutual
Information (SPPMI) matrix derived from feature co-occurrence data at site $k$
\citep{levy2014neural, li2024multisource}, which provides a widely used
representation of feature similarity based on co-occurrence patterns.

We model $\blS_k$ using a low-rank signal plus noise
decomposition,
\[
\blS_k = \blM_k + \blE_k, \quad
\blM_k = \blX_k \blX_k^\top = \blU_k \blD_k \blU_k^\top, 
\]
for $k=1,2$,
where $\blM_k \in \bbR^{n_k \times n_k}$ is a symmetric rank-$r$ signal matrix and
$\blE_k \in \bbR^{n_k \times n_k}$ is a mean-zero symmetric noise matrix with
independent upper-triangular entries. The diagonal matrix
$\blD_k \in \bbR^{r \times r}$ contains the nonzero eigenvalues of $\blM_k$, and
$\blU_k \in \bbR^{n_k \times r}$ contains the associated eigenvectors. We define
$\blX_k = \blU_k \blD_k^{1/2} =( \blx_{k,1},\dots \blx_{k, n_k})\trans \in \bbR^{n_k \times r}$
as the population-level embedding matrix for the features observed at site $k$, and $\blx_{k,i}\in\bbR^{ r}$ denotes the true embedding of feature $i$ at site $k$.
This formulation is motivated by the fact that the SVD
of SPPMI matrices recovers skip-gram-with-negative-sampling embeddings
\citep{levy2014neural}, providing a principled interpretation of $\blX_k$ as the
latent feature embeddings at site $k$.

In addition to the site-specific matrices $\blS_1$ and $\blS_2$, we leverage an external grouping structure shared across sites. Let $G$ denote the total number of groups, and let $g_i \in [G]$ indicate the group membership of feature $i$. This hierarchical structure provides valuable prior information: features within the same group are expected to have similar embeddings, up to a small number of group-specific outliers whose representations deviate substantially from the group pattern.
For example, in EHR data, Ibuprofen and Naproxen are both nonsteroidal anti-inflammatory drugs and thus play similar clinical roles, suggesting their embeddings should exhibit high similarity. By incorporating this group information, we aim to improve embedding estimation efficiency, enhance cross-site alignment, and stabilize embedding recovery in the presence of multi-block missingness.

Let $\cV = \cV_1 \cup \cV_2$,  $\cV_o=\cV_1\cap\cV_2$ with $|\cV_o| = n_o$, and $\cV_{no}=\cV \setminus \cV_{o}$ denote the index sets of all, overlapping, and non-overlapping features, respectively.
We assume some overlapping features admit a shared population embedding across the two systems. We refer to these as \emph{cross-site consistent}
features and denote the set by $\sot$.
Others exhibit genuine cross-site heterogeneity and admit distinct population embeddings across sites. We refer to these as \emph{cross-site divergent}
features and denote the set by $\cV_{o,div}$.
For each group, we assume most features share a similar embedding (excluding cross-site divergent features) and allow the existence of group outliers whose embeddings deviate
significantly from the majority in that group. 
We denote by $\mathcal{V}_{out}$ the set of features whose population embeddings
do not conform to either cross-site or within-group homogeneity assumptions,
including cross-site divergent features and group outliers.
The remaining features form
$\cV_{trans}=\sot\cup\cV_{no,grp}$, where $\cV_{no,grp}$ contains
non-overlapping features that are not group outliers. We further divide
$\cV_{no,grp}$ into a set of features belonging to groups anchored by at least one
cross-site consistent overlapping code, denoted by $\sonet$, and a set of features in
groups without any cross-site consistent features, denoted by $\stwot$. These four sets,  $\{\sot, \sonet, \stwot, \sthreet\}$, 
determine how information is shared or
restricted across sites and groups.

For any feature $i\in \cV\setminus\cV_{o,div}$,  we introduce an indicator $h_i$, where $h_i = 1$ if the feature is not a group outlier and $h_i = 0$ otherwise.
We assume the existence of a population-level partition
$\mathcal P = \{ P_1, \ldots, P_L \}$
of $\mathcal V_{o,cons}$ such that features within the same block share a common population embedding. 
Each block corresponds to either a group shared across sites or a singleton
feature whose embedding is consistent across sites. Formally, for any $i,i' \in P_\ell$ with $|P_\ell|>1$, we have $g_i = g_{i'}$, $h_i = h_{i'} =1$, and 
\[
\blx_{1,i} = \blx_{1,i'} = \blx_{2,i} = \blx_{2,i'}.
\]
In addition, we impose population-level consistency for site-specific features through their group structure.
For any site-specific feature $l \in \mathcal{V}_{grp,anch}
\cap \cV_k$, there exists at least one feature $j \in \sot$ such that $g_l = g_j$ and $h_l = h_j = 1$. We assume
$\blx_{k,l} = \blx_{1,j} = \blx_{2,j}$.
Finally, for site-specific features $i_1 \in \stwot \cap \cV_1$ and $i_2 \in \stwot\cap \cV_2$ with $g_{i_1} = g_{i_2}$ and $h_{i_1} = h_{i_2} = 1$, 
belonging to the same group containing no cross-site consistent features, we assume
$\blx_{1,i_1} = \blx_{2,i_2}$.

\begin{figure}[b]
    \centering
    \includegraphics[width=1\linewidth]{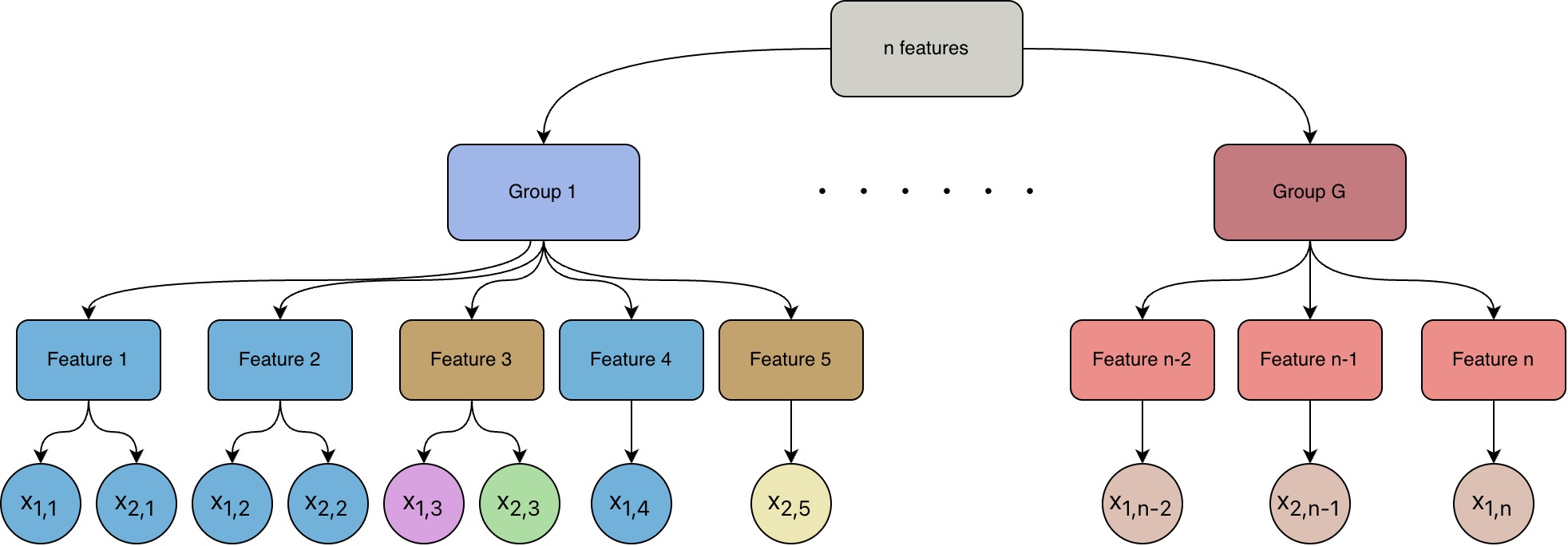}
    \caption{Illustration of the hierarchical structure of $n$ features: features 1 and 2 belong to $\sot$, features 3 and 5 belong to $\sthreet$, feature 4 belongs to $\sonet$, and features $n-2$, $n-1$ and $n$ belong to $\stwot$.  
}
    \label{fig:hierarchical}
\end{figure}

Figure~\ref{fig:hierarchical} illustrates the hierarchical feature structure and
the corresponding population-level embedding relationships across sites.
The estimation procedure described in the next section is designed to recover
this hierarchical structure by identifying feature types and enforcing the
corresponding embedding constraints.

\subsection{Overview of the Estimation Method}

The goal of our estimation procedure is to co-train embeddings across two sites
while selectively borrowing information for features estimated to belong to
$\mathcal{V}_{trans}$ and avoiding negative transfer for those estimated to
belong to $\mathcal{V}_{out}$.
To this end, the algorithm combines initial single-site embeddings with a
hierarchical feature classification and then refines the embeddings according to the resulting feature types.

\begin{figure}[b]
    \centering
    \includegraphics[width=1\linewidth]{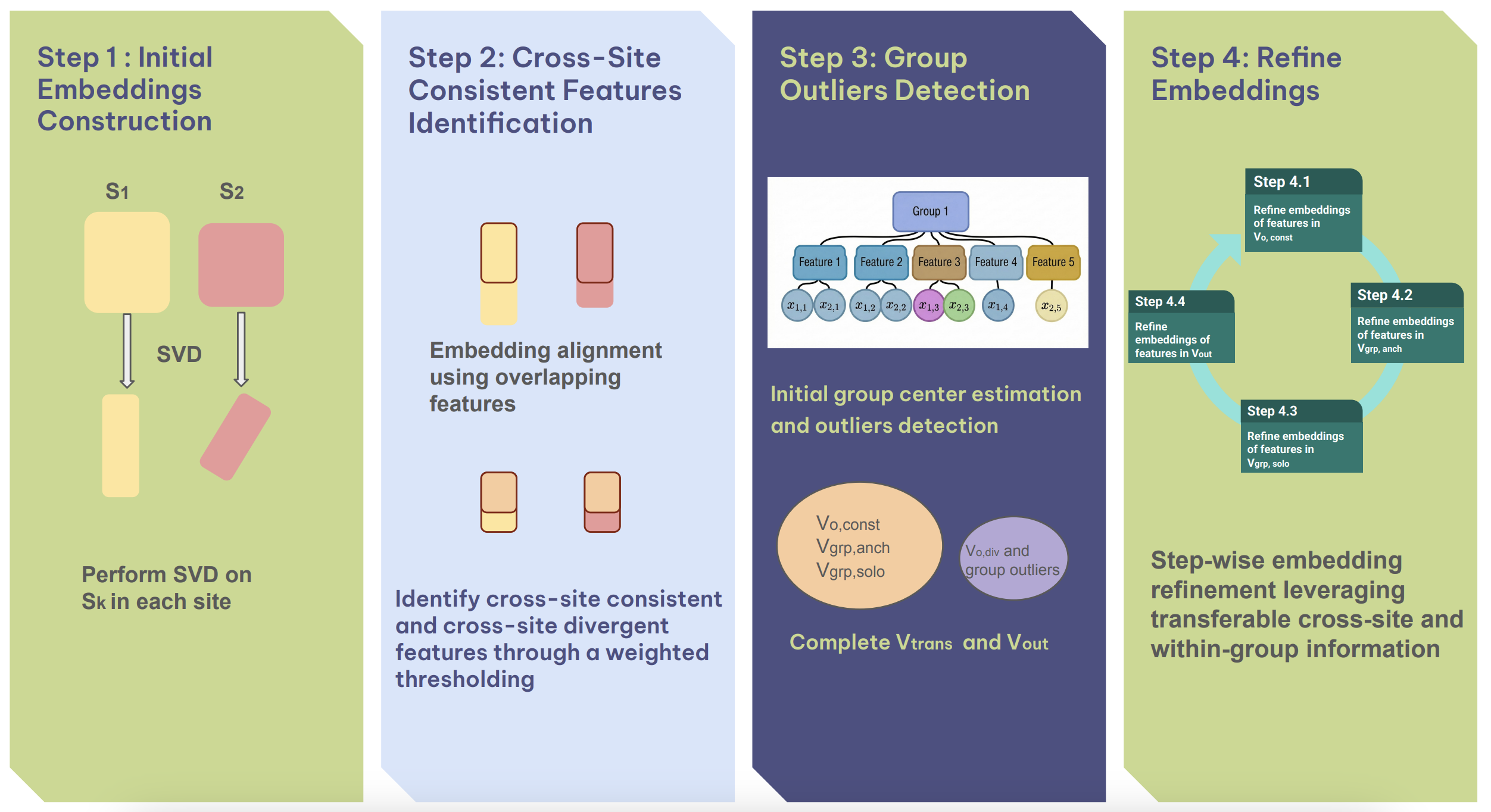}
    \caption{Schematic of the TransNEST algorithm.}

    \label{fig:Schematic}
\end{figure}

As illustrated in Figure~\ref{fig:Schematic}, the algorithm proceeds in
four main steps:
\begin{enumerate}
    \item Obtain initial embeddings within each site via SVD of
    $\blS_k$.
    \item Identify cross-site consistent features via a weighted
    thresholding rule on aligned initial embeddings, 
    thereby constructing estimates $\so$ and the overlapping portion of $\sthree$.

    \item Compute a provisional group center (excluding
    cross-site divergent features) for each group and use a second thresholding rule to detect group outliers whose embeddings deviate significantly from this center, 
    completing the construction of $\sthree$, $\sone$, and $\stwo$.
    
    \item Refine embeddings for features in $\so$, $\sone$, $\stwo$, and
    $\sthree$ in a stepwise fashion, leveraging cross-site and
    within-group information where appropriate, and treating heterogeneous features separately.
\end{enumerate}

\begin{remark}
The proposed framework is not limited to settings in which sites share identical
or directly overlapping feature sets. More generally, it applies whenever a subset of features admits known one-to-one correspondences across sites, while the remaining features are site-specific and organized by shared group
structure.
For example, in multilingual representation learning, site 1 and site 2 may correspond to
corpora in different languages, such as English and Haitian, with partially overlapping vocabularies or known
cross-lingual alignments. In this case, $\blS_k$ represents language-specific
SPPMI matrices, aligned words are linked through bilingual dictionaries, and $g_i$ encodes shared linguistic categories such as part of speech or semantic class.
TransNEST can then refine Haitian word embeddings by transferring co-occurrence structure learned from the information rich English corpus and the shared group structure.

\end{remark}

\subsection{Detailed Estimation Method}\label{subsubsec:alg}

\subsubsection*{Step 1: Initial Single-Site Embeddings via SVD.}

We obtain initial embeddings by performing an SVD of $\blS_k$ and
approximating it with the rank-$r$ matrix
$\hat\blS_k=\hat{\blU}_k \hat{\blD}_k \hat{\blU}_k^\top$, where
$\hat{\blD}_k\in\bbR^{r\times r}$ contains the top $r$ singular values
(in descending order) and $\hat{\blU}_k\in\bbR^{n_k\times r}$ contains
the corresponding left singular vectors. The initial embedding matrix is
then
\[
\embSVD_k=\hat{\blU}_k \hat{\blD}_k^{1/2}\in\bbR^{n_k\times r},
\]
and we denote by $\embsvd_{k,i}$ the $i$th row of $\embSVD_k$, which
serves as the initial embedding for feature $i$ at site $k$.

\begin{remark}
Although we use single-site SVD embeddings as the initial estimator in this study, other consistent embeddings, such as those derived from large language models (LLMs) \citep[e.g.][]{rasmy2021med, yuan2022coder}, can also  be integrated to initialize our algorithm. The error rate of the refined TransNEST estimator can be derived using our proof framework, incorporating the initial estimation error of the chosen method.
\end{remark}

\subsubsection*{Step 2: Embedding Alignment and Identification of Cross-Site
Consistent Features.}

Because embeddings are identifiable only up to rotation, we first align
the initial single-site embeddings. Specifically, for any orthogonal matrix
$\blQ\in\Orth$, $\embSVD_k$ and $\embSVD_k\blQ$ represent the same
solution. To make cross-site comparisons meaningful, we rotate the
embeddings at site 2 to best match those at site 1.

Let $\embSVD_{k,o}$ denote the submatrix of $\embSVD_k$ for overlapping
features. We compute the optimal rotation
\[
\hat{\blQ}
=\mathbb{W}(\embSVD_{2,o},\embSVD_{1,o})
=\arg\min_{\blQ\in\Orth}
\|\embSVD_{1,o}-\embSVD_{2,o}\blQ\|_F,
\]
and define the aligned embeddings as
$\embSVDa_1=\embSVD_1$,
$\embSVDa_2=\embSVD_2 \hat{\blQ}$.
Let $\embsvda_{k,i}$ denote the $i$th row of $\embSVDa_k$, $k=1,2$.

Using the aligned embeddings, we classify each overlapping feature as
either \emph{cross-site consistent} or \emph{cross-site divergent}. A
feature is deemed cross-site consistent if its embeddings at the two sites
are sufficiently similar:
\[
\so
=\{\,i\in\cV_o:\|\embsvda_{1,i}-\embsvda_{2,i}\|\le\lambda_i\,\},
\]
and cross-site divergent otherwise:
$\widehat{\cV}_{o, div}=\cV_o\setminus\so$,
where $\lambda_i$ is a feature-specific threshold.

\begin{remark}
To simplify tuning, we incorporate site-specific feature weight $w_{k,i}$,
e.g., marginal frequency of feature $i$ at site $k$. Define the weighted mean embedding
\[
\embsvda_i=\frac{w_{1,i}\embsvda_{1,i}+w_{2,i}\embsvda_{2,i}}
{w_{1,i}+w_{2,i}},
\]
and classify $i\in\cV_o$ as cross-site consistent if
\begin{equation*}
\begin{split}
\max\Big\{
w_{1,i}(w_{1,i}+w_{2,i})\|\embsvda_{1,i}-\embsvda_i\|^2,\; \\
w_{2,i}(w_{1,i}+w_{2,i})\|\embsvda_{2,i}-\embsvda_i\|^2
\Big\}
\le\lambda.     
\end{split}
\end{equation*}
In this formulation, $\lambda_i$ reduces to a closed-form expression
depending on $(w_{1,i},w_{2,i},\lambda)$, so only one hyperparameter
$\lambda$ needs to be tuned. This design is naturally more tolerant of estimation variability for infrequent
features, encouraging knowledge transfer toward them. 
\end{remark}

\subsubsection*{Step 3: Initial Group Centers Estimation and a Thresholding Rule for Merging Blocks.}

We obtain initial estimates for group centers and use a thresholding rule to decide whether each feature should be merged into its group center. 
We propose to estimate each group center as the weighted mean of features in the group, excluding any cross-site divergent features. We use $\gginithat$ to denote the initial group center of group $g \in [G]$, and it is calculated as 
$$
\gginithat = \frac{\sum_{i} \sum_{k=1}^2 w_{k, i}\embsvda_{k, i}\mathbb{I}\{g_i= g, i \in \cV_k\setminus \widehat{\cV}_{o, div}\}}{ w_g}, 
$$	
where $w_g = \sum_{i} \sum_{k=1}^2 w_{k, i}\mathbb{I}\{g_i = g,i \in \cV_k\setminus\widehat{\cV}_{o, div}\}$. If the feature-site weights $w_{k, i}$ are unavailable, we can set all of them to one. 

The thresholding rule is defined as follows. 
Feature $i$ is merged into the center of group $g_i$ if and only if
$\hat h_i = 1$, where
$$
\hat{h}_i = \mathbb{I} \left(\max_{k=1, 2}\big[ c_{k, i}\|\embsvda_{k, i}-\ginithat_{g_i}\|^2 \mathbb{I}\{i\in\cV_k\setminus\widehat{\cV}_{o, div}\} \big] \leq \mu_i\right).
$$
Here $c_{k,i}$ are pre-specified weights, $\mu_i$ are tuning parameters, and the
indicator restricts the decision to observed, non-divergent features. Thus,
$\hat h_i = 1$ indicates that feature $i$ is treated as group-homogeneous and
merged into its group center, while $\hat h_i = 0$ indicates a group outlier.

\begin{remark}
In practice, when feature-site weights are available, one possible choice of $c_{k, i}$ can be  $w_{k, i} w_{g_i}$.
We can further set $\mu_i = \mu$, the only hyperparameter that needs to be tuned.  
We adopt this configuration in the numerical studies.
Similarly, this configuration is more tolerant of within-group heterogeneity for groups with smaller weights or that are less frequent. 
\end{remark}

\subsubsection*{Step 4: Embedding Refinement.}
After applying the two thresholding rules, 
we obtain the estimated four sets:
$\so$,
\begin{equation*}
\begin{split}
  \sone:=\{i \in \cV_{no}: \hat{h}_i=1 \text{ and } \exists \, i'\in\so \\ \text{ such that } \hat{h}_{i'}=1, g_i = g_{i^\prime} \},  
\end{split}
\end{equation*}
\begin{equation*}
\begin{split}
  \stwo:=\{i \in \cV_{no}: \hat{h}_i=1 \text{ and } \nexists \,i'\in \so \\ \text{ such that } \hat{h}_{i'}=1,  g_i = g_{i^\prime} \},  
\end{split}
\end{equation*}
$$ \text{and} \quad \sthree:=\big(\cV_{no}\cap\{i: \hat{h}_i=0\}\big)\cup \widehat{\cV}_{o, div}.$$ 
We then refine the embeddings for features in each set using update rules tailored to their estimated cross-site and within-group transferability.

Step 4.1: Refine embeddings of features in $\so$.
We refine embeddings for features in $\widehat{\mathcal V}_{o,cons}$ by enforcing
the block structure learned in steps 1-3, denoted by $\widehat{\mathcal P} = \{ \widehat P_1, \ldots, \widehat P_{\widehat L} \}$. 
For any $i$ and $i' \in \widehat P_l$, $g_i = g_{i'}$ and $\hat h_i = \hat h_{i'} =1$.
Let
$$\blX_{k,\subso} = (\blx_{k,i})_{i\in \subso} \in \bbR^{|\widehat{\mathcal V}_{o,cons}| \times r}$$
denote the population embedding matrix at site $k$, and let
$\mathbf S_{k,\widehat{\mathcal V}_{o,cons}}$ be the corresponding submatrix of
$\mathbf S_k$.
Within each block $\widehat P_\ell$, embeddings are constrained to be identical
across sites.
Hence, we define the constrained parameter space
\begin{equation*}
    \begin{split}
       \mathcal{M}_{\hat{\blh}}=\{\blX_{1,\subso}, \blX_{2,\subso}: \blx_{1,i} =  \blx_{1,i'}  = \blx_{2,i} = \blx_{2,i'}, \\ \forall i, i' \in \hat P_l, l\in[\hat L]\}.   
    \end{split}
\end{equation*}
The refined embeddings are obtained by solving
\begin{equation}\label{eq:cotrain}
\begin{split}
(\hat{\blX}_{1,\subso},\hat{\blX}_{2,\subso})
=\arg\min_{(\blX_{1,\subso},\blX_{2,\subso})\in\mathcal{M}_{\hat{\blh}}} \\
\sum_{k=1,2}\tilde w_k\,
\bigl\|\blS_{k,\subso}-\blX_{k,\subso}\blX_{k,\subso}^{\top}\bigr\|_F^{2},
\end{split}
\end{equation}
where $\tilde{w}_k$ is the weight of site $k$.
For any feature $i \in \so$ , we denote its refined embedding as $\hat \blx_{1,i} = \hat \blx_{2,i}$.

\begin{remark}
The weight of site  $\tilde{w}_{k}$ plays an important role in (\ref{eq:cotrain}) and potentially improving the estimation efficiency.
For example, if site $1$ has much less noise in the construction of $\blS_{1}$ (e.g., hospital $1$ has much more patients using to construct the co-occurrence matrix of EHR features) compared with site $2$, then  $\tilde{w}_{1}$ should be much larger than  $\tilde{w}_{2}$. 
In practice, these site weights can be obtained from data \citep{zhou2021multi}, external information (e.g., number of patients from each hospital), or through a tuning procedure. 
Unlike steps 2 and 3, where feature-site weights $w_{k,i}$ are used to enhance accurate feature type detection,  we employ the site weight $\tilde{w}_k$ here to simplify the computation.
\end{remark}

Step 4.2: Refine embeddings of features in $\sone$.
For $i\in \sone$, there is a feature $i'\in \so$ such that they belong to one group, i.e., $g_{i'}=g_{i}$, and both are similar to the group center, i.e., $\hat{h}_{i'}=\hat{h}_{i}=1$. We directly set $\hat{\blx}_{k,i} =  \hat{\blx}_{k, i'}$,  where $\hat{\blx}_{k, i'}$ is obtained in Step 4.1.

Step 4.3: Refine embeddings of features in $\stwo$.
For non-overlapping features similar to the center of the group consisting of only non-overlapping features, we propose to refine their embeddings by conducting a weighted linear regression, treating refined $\hat{\blX}_{k, \subso}$ as the design matrices. 
For each $g\in\{g_i:i\in \stwo\}$, we estimate the shared group embedding as
\begin{equation*}
\begin{split}
    \hat{\bggamma}_{g} = \arg\min_{\bggamma}\sum_{k=1,2}\tilde{w}_k\sum_{i}\sum_{j}\Big\{(s_{k,i,j} -\bggamma^T\hat{\blx}_{k,j} )^2  \\\times \mathbb{I}(i\in \cV_k \cap\stwo, g_i = g, j \in\so) \Big\}.
\end{split}
\end{equation*}
We apply the site weight $\tilde{w}_k$ again to potentially enhance estimation efficiency. 
For any feature $i \in \stwo$, we denote its refined embedding as $\hat \blx_{k,i} = \hat{\bggamma}_{g_i}$.

Step 4.4: Refine embeddings of features in $\sthree$.
For features in $\sthree$, where direct knowledge transfer is impossible, we refine their embeddings by solving linear regressions at each site using the refined embeddings of features in $\widehat \cV_{trans} =  \so\cup\sone\cup\stwo$, together with the initial single-site SVD embeddings of features in $\sthree$, as the full design matrices. 

Before the regression, we first need to align the refined embeddings and the initial single-site embeddings at each site. 
Specifically, we estimate the orthogonal matrix $\blW_{k,r}$ at each site by aligning  $\embsvd_{k, i}$ with $\hat{\blx}_{k, i}$ for $i\in \widehat \cV_{trans}$, and 
$\blW_{k,r} =\mathbb{W}\{\embSVD_{k, \subtrans}, \hat{\blX}_{k, \subtrans}\}$, where $\embSVD_{k,\subtrans} = (\tilde\blx_{k,i})_{i\in \subtrans}$ and $\hat{\blX}_{k, \subtrans} = (\hat\blx_{k,i})_{i\in \subtrans}$.
Let $\embsvdaa_{k, i} = \blW_{k,r}^T\embsvd_{k, i}$, which is aligned with the refined embeddings at site $k$. We then solve for $\hat{\blx}_{k, i}$, $i\in \sthree$  as follows 
\begin{equation}\label{eq:step4.4}
\begin{split}
	\hat{\blx}_{k, i}=\arg\min_{\blx}\Big\{\sum_{j}(s_{k,i,j} -\blx^T\hat{\blx}_{k,j} )^2 \mathbb{I}(j\in\cV_k \setminus \sthree) \\ + \sum_{j}(s_{k,i,j} -\blx^T\embsvdaa_{k,j} )^2\mathbb{I} (j\in\sthree)\Big\}.
\end{split}
\end{equation}

\begin{remark}
Compared with the initial single-site embedding estimation, 
here we use the refined embeddings as part of the design matrix, which can be interpreted as reducing measurement error in covariates, resulting in better empirical performance.
The term $\sum_{j}(s_{k,i,j} -\blx^T\embsvdaa_{k,j} )^2\mathbb{I} (j\in\sthree)$ in the loss function is used to guarantee that our embeddings are no worse than the initial single-site embeddings in the worst scenario where $\widehat\cV_{trans} = \emptyset$. 
\end{remark}

\section{Theoretical Results}

\subsection{Assumptions}
In this subsection, we introduce the most important Assumption \ref{ass:refine} on the relationship among the number of different types of features and the order of the largest and the $r$-th largest singular values of the true embedding matrices, $\blX_k$ and $\blX_{k,\subsot}$, $k=1,2$.
We defer other more technical assumptions, Assumptions S1-- S4,
on the sub-Gaussian error $\blE_k$ with $\sigma_k^2$ as the sub-Gaussian parameter, the incoherence condition of $\blU_k$, the large enough gap between the two embeddings of any cross-site divergent features and between the outliers of a group and the group center for selection consistency, to Section A in the Supplementary Materials.
To simplify the error rates, we assume the embedding rank $r$ is fixed. Detailed error rates as functions of $r$ are provided in the Supplementary Materials.

\begin{assumption}\label{ass:refine}
We assume the following conditions hold:
\begin{enumerate}
    \item[(i)] The numbers of features at the two sites and in their overlap are of the same order, that is,
    $n_1 \sim n_2 \sim n_o \sim n$.
    
    \item[(ii)] The leading and $r$-th  singular values of the true embedding matrices at both sites, as well as those associated with the cross-site consistent features, are of the same order:
    \begin{equation*}
    \begin{split}
    \tau_1^2(\blX_1) \sim \tau_1^2(\blX_2) \sim \tau_r(\blX_{2,\subsot}^T \blX_{1,\subsot}) \\
    \sim \tau_r^2(\blX_{1,\subsot}) \sim \tau_r^2(\blX_{2,\subsot}).       
    \end{split}
    \end{equation*}
    This implies that the number of population-level blocks satisfies $|\mathcal{P}| \ge r$.
    
    \item[(iii)] The signal strength of the embeddings dominates the noise level, in the sense that $\tau_r^2(\blX_k) \gtrsim n_k \sigma_k$, for $k = 1,2$.

    \item[(iv)] The effective complexity of the non-transferable feature set is controlled, with $ |\sthreet| \lesssim r^{-1} n$.
\end{enumerate}
\end{assumption}

\begin{remark} 
Condition (i) of Assumption \ref{ass:refine} requires a sufficient number of overlapping features to enable effective information transfer across sites. Conditions (ii)–(iii) impose a signal strength requirement ensuring that the shared latent structure is identifiable in the presence of noise. Condition (iv) controls the size of cross-site divergent and group-outlying features, preventing them from overwhelming the shared structure.
\end{remark}

\subsection{Selection Consistency}

In this subsection, we show model selection consistency in terms of identifying cross-site consistent features and merging blocks based on the prior group structures in the following two Lemmas. The detailed proofs are provided in Sections E and G of the Supplement Materials. 
\begin{lemma}\label{lemma:so-hat-so}
    Under Assumptions \ref{ass:refine} and  S1 -- S4, we can consistently identify cross-site consistent features via the proposed thresholding procedure in step 2, i.e., 
    \begin{equation}
		\lim_{n\to\infty}\pr(\so =  \sot) = 1.
	\end{equation}
\end{lemma}

\begin{lemma}\label{lemma:MergeBlocksConsistency}
Under Assumptions  \ref{ass:refine} and  S1 -- S4, for all features except cross-site divergent features, 
we can consistently merge blocks via the proposed thresholding procedure in step 3, i.e., 
\begin{equation}
\lim_{n\to\infty}\pr\Big(\hat{h}_i=h_i, \text{ for all } i \in \cV \setminus \cV_{o,div}\Big)=1.
\end{equation}
\end{lemma}

\subsection{Error Bounds after Refinements }
We establish error bounds for feature embedding refinements on the sets 
$\sot$, $\sonet$, $\stwot$, and $\sthreet$ 
in Propositions S1 -- S4, respectively, in the Supplementary Materials. 
In the main paper, we present only the final error rate for $\min_{\blW_X \in \mathcal O^r}\|\hat{\blX}_k - \blX_k \blW_X\|_F$,
which is obtained by aggregating the results of the four propositions.

To provide a simplified and interpretable error bound, we introduce Assumption S5, under which we assume that the embedding rank $r$ is fixed, population-level groups are not severely underrepresented at any site, and site-specific features are balanced across sites.
The full error bound without Assumption S5 is given in 
Theorem S1 in the Supplementary Materials.

We summarize the resulting simplified error rate of the TransNEST estimator in the following theorem.
\begin{theorem}\label{th1}
Under Assumptions \ref{ass:refine} and  S1 -- S5, if $\tilde{w}_1/\tilde{w}_2\sim \sigma_2^2/\sigma_1^2$, for $k=1, 2$, we have  
\begin{equation}
\begin{split}
\min_{\blW_{X}\in \mathcal O^r}\|\hat{\blX}_k-\blX_k\blW_X\|_F 
 = O_p \Big[ \log^{1/2}(n) \tau_r^{-1}(\blX_{k}) \\
 \times\big\{(|\cP|^{1/2} + |G_{solo}|^{1/2})(\sigma_1\wedge\sigma_2)  +|\sthreet|^{1/2}\sigma_k \big\} \Big],
\end{split}
\end{equation}
where $G_{solo}= \{g_i: i\in \stwot\}$.
\end{theorem}
Here $\blW_X \in \mathcal O^r$ denotes an orthogonal matrix achieving the optimal Procrustes alignment between $\hat{\blX}_k$ and $\blX_k$.
The presence of group structure and cross-site knowledge transfer reduces the effective number of parameters from $(n_1+n_2)r$ to $(|\mathcal P|+|G_{\mathrm{solo}}|+|\sthreet|)r$. 
With appropriately chosen site weights, the effective noise level for features in $\cV_{trans}$ is reduced from $\sigma_k$ to $\sigma_1\wedge\sigma_2$. 
Without loss of generality, we treat site~2 as the target site with $\sigma_1<\sigma_2$.
Compared with a target-only SVD estimator, whose error rate is of the order of 
$\tau_r^{-1}(\blX_2)n_2^{1/2}\sigma_2$, 
the TransNEST estimator achieves a strictly better rate when
\[
(|\cP|^{1/2}+|G_{solo}|^{1/2})\sigma_1 \ll n_2^{1/2}\sigma_2
\quad \text{and} \quad
|\sthreet| \ll n.
\]

\section{Simulation Studies}
For data generation, we consider the following embedding structure:
\[
{\blx}_{k,i}:= \bbbb_{g_i}\mathbb{I}(|G_i|>1) + \zzz_{i} + \ddd_{k,i},
\]
where $\bbbb_{g_i}$ is the embedding of the group effect, 
$|G_i|$ is the number of features in group $g_i$,
$\zzz_{i}$ is the embedding of the feature effect shared between sites, and  $\ddd_{k,i}$ is the embedding of the feature-site effect. 
If feature $i$ belongs to $\sot$, $\ddd_{1,i} = \ddd_{2,i} = \bm 0$.
If feature $i$ at site $k$ has the same embedding as its group center, then $ \zzz_{i} + \ddd_{k,i} = \bm 0$. 
We set $n_1 = n_2 = 2000$, $n = 3000$, $n_o = 1000$, and $r=50$.

Let $\BB = (\bbb_1, \dots, \bbb_G)\trans \in \mathbb{R}^{G \times r}$. We generate group effects via a multivariate normal distribution, i.e., 
$\BB_{\cdot, j}\sim N(0, \boldsymbol{\Sigma}_1)$, $j\in [r]$, where $\boldsymbol{\Sigma}_1 \in\mathbb{R}^{G\times G}$ is block diagonal.
Specifically, $\boldsymbol{\Sigma}_1$ consists of $G/10$ repeated non-overlapping blocks, denoted by $\DD \in \mathbb{R}^{10\times 10}$.
$\DD$ is also block diagonal, with two sub-blocks. 
The first sub-block is a $3\times3$ autoregressive covariance matrix of order 1 (AR1) with correlation parameter $\rho_\beta = 0.4$, i.e., 
$D_{ij} = 0.4^{|i-j|}$ for $i,j \in[3]$.
This implies that the corresponding groups are related.
The second sub-block is a $7\times7$ identity matrix. 
Further, to mimic the hierarchical ontology structure of EHR features in Section 5, 500 features from each site do not belong to any groups.
Let $\ZZ = (\zzz_1, \dots, \zzz_n)\trans \in \mathbb{R}^{n \times r}$. 
We first generate feature effects via a multivariate normal distribution, i.e., 
$\ZZ_{\cdot, j} \sim N(0, \boldsymbol{\Sigma}_2)$, $j\in [r]$,  
where $\boldsymbol{\Sigma}_2 \in\mathbb{R}^{n\times n}$ is block diagonal, and each block is AR1 with size $6$ and  correlation parameter  $\rho_\zeta = 0.4$.
We then randomly select $n_\zeta^c$ overlapping features, and set their $\zzz_i$ to zero. For these features, the group effect dominates.
Regarding feature-site effects, we randomly select $n_\delta$ overlapping features from each site, denoting the set of indexes as $\{d_{k,1}, \dots, d_{k, n_\delta}\}$, $k=1, 2$, and generate feature-site effects for them. While for the remaining features, $\ddd_{k,i} = \bm 0$. 
Let  $\DDDD_k = (\ddd_{k,d_{k,1}}, \dots, \ddd_{k,d_{k,n_\delta}})\trans \in \mathbb{R}^{n_\delta \times r}$.
We generate feature-site effects via a multivariate normal distribution, i.e., 
$\DDDD_{k, \cdot, j}\sim N(0, \boldsymbol{\Sigma}_3)$, $j\in[r]$, $k=1, 2$,
where $\boldsymbol{\Sigma}_3 \in\mathbb{R}^{ n_\delta\times n_\delta}$ is block diagonal with block size 50 and each block is AR1 with correlation parameter $\rho_\delta =0.95$.

We then generate the data matrices $\SSS_k$ via 
$s_{k, i, j} = \blx_{k,i}\trans \blx_{k,i}+ \epsilon_{k, i, j}$, $ \epsilon_{ij} \sim N(0, \sigma_{k,i}\sigma_{k,j})$.
We treat site 1 as the source site with $\sigma_{1,i} = 5$ for any $i\in [n_1]$. 
For the target site 2, we further separate features into frequent features and rare features to mimic the real EHR features. 
Specifically, there are $n_f = 1300$ frequent features with $\sigma_{2, f} = 20$, and $n_r = 700$ rare features with $\sigma_{2, r} = 80$.
We set $w_{k, i}= 1/\sigma_{k,i}$ in the simulation, and similar to \cite{zhou2023multi}, we calculate the site weights as 
\begin{equation}\label{eq:site_weight}
\widetilde{w}_{k} = \frac{n_{3-k}^{-1} \|\blS_{3-k}  - \embSVD_{3-k}(\embSVD_{3-k})^T\|_2^2}{n_1^{-1} \|\blS_1  - \embSVD_1(\embSVD_1)^T\|_2^2+n_2^{-1} \|\blS_2  - \embSVD_2(\embSVD_2)^T\|_2^2}.     
\end{equation}

We compare our method with four benchmarks: (1) single-site SVD (SSVD), i.e., $\embSVD_2$,  (2) single-site with group structures (SSG) method, (3) plain data pooling (DP) method, and (4) BONMI method \citep{zhou2023multi}.
The SSVD method fails to leverage the group structures and the information-rich source population.
For the SSG method, we utilize the prior group structures and replace the feature embedding with the mean of the SSVD embeddings of all features belonging to the group. This single-site method ignores the heterogeneity within groups. 
In the DP method, we first calculate the weighted sum of the observed data matrices for the overlapping features, i.e.,  $\sum_{k=1,2}\widetilde w_k \{S_{k, i, j}\}_{i, j \in \cV_o}$, and obtain the embedding of overlapping features via SVD; we then solve an orthogonal procrustes problem using the co-trained embeddings and the SSVD embeddings of overlapping features at the target site; we finally rotate the SSVD embedding of target-specific features using the estimated orthogonal matrix to get aligned DP embeddings.
BONMI method completes the full matrix $\SSS_{\rm full} \in \mathbb{R}^{n\times n}$ by recovering the missing block $\SSS_{\rm miss} \in \mathbb{R}^{(n1- n_o)\times (n2 - n_o)}$, i.e., the relationship between site-1-specific features and site-2-specific features, under the low-rank assumption, and then applies SVD to the completed $\widehat\SSS_{\rm full}$ to obtain the embeddings for all $n$ features. 
Note that DP and BONMI both ignore the heterogeneity between sites and do not leverage the prior group structures.

We consider configurations where $G=250, 400$, $n_\zeta^c = 0, 1000$, and $n_\delta = 100, 200$ to study the influence of group size, 
the level of group-heterogeneity, and the level of site-heterogeneity on the performance of different methods. When $G = 250$, the uniform group size is 8; when $G = 400$, the uniform group size is 5.
When $n_\zeta^c = 0$, then Assumption S4 on the exact group selection consistency does not hold. We explore the performance of TransNEST under this model misspecification.

For evaluation, we not only report the estimation accuracy in terms of estimating the true embedding $\blX_2$, but also the performance of identifying positive feature-feature pairs against random pairs, motivated by EHR applications. 
In practice, we have knowledge of which two EHR features are similar and which two are related. For example, type 1 diabetes is similar to type 2 diabetes, and they belong to the same group, diabetes mellitus. 
Type 2 diabetes is related to insulin, although they do not belong to the same ontology group. 
Further, the sparsity of the network allows us to assume that each random pair represents a weak or no relationship.
We curate related feature-feature pairs based on the structure of $\boldsymbol{\Sigma}_l$, $l=1, 2, 3$, for evaluation. 
If two features belong to the same group, they are similar and related.
If ${\Sigma_{1, i,j}} > 0$, then features in the $i$th group are related to the features in the $j$th group. 
If ${\Sigma_{2, i,j}} > 0$, $\zzz_i \neq \bm 0$ and  $\zzz_j \neq \bm 0$, then feature $i$ and feature $j$ are related.
We can further expand the set of related pairs specific for the target site using the non-zero elements in $\boldsymbol{\Sigma}_3$ paired with the set $\{d_{2,1}, \dots, d_{2, n_\delta}\}$. These target-specific pairs mimic the EHR feature pairs that hold only for a specific patient sub-population.
A random subset of the positive pairs was used for tuning, and the remaining set was used for evaluation. 

The results of the four configurations with $G = 400$ are summarized in Tables \ref{table1} and \ref{table2}. 
In Table~\ref{table1}, we report area under the curve (AUC) for all available feature pairs, frequent pairs, rare pairs, frequent and transfer-eligible pairs, frequent and transfer-ineligible pairs, rare and transfer-eligible pairs, and rare and transfer-ineligible pairs.
If one of the features in a pair is frequent and allows knowledge transfer, i.e., belonging to $\cV_{trans}$, we classify the pair as frequent and transfer eligible. The same classification logic is applied to the other types of feature pairs.
For each positive or random feature pair in the evaluation set, we compute the cosine similarity between the embeddings of the two features. AUCs are then calculated by comparing the vector of cosine similarities to a binary vector of the same length, where 1 indicates a positive pair, and 0 indicates a random (negative) pair.
Our method, TransNEST, achieves the highest overall AUC for distinguishing positive pairs from random pairs across all four configurations.
Relative to $(C1)$, configuration $(C3)$ exhibits greater heterogeneity within groups, which impairs the performance of SSG and reduces the effective sample size available to TransNEST. Despite this, TransNEST generally outperforms the other methods by adaptively capturing the degree of heterogeneity and borrowing useful information from the source.
Compared with $(C1)$, $(C2)$ has a larger $n_\delta$,  
indicating a higher level of heterogeneity both between sites and within groups. This increased heterogeneity degrades the performance of SSG, DP, and BONMI; however, SSVD and TransNEST remain relatively robust. 
For TransNEST, the observed decrease in AUCs for rare and transfer-ineligible features is primarily due to the smaller number of cross-site consistent features and the smaller effective sample sizes when updating their embeddings. 
As a result, the data matrix used for updating the embeddings of rare and transfer-ineligible features, via the regression framework, is of lower quality, negatively affecting performance.
Configuration $(C4)$ presents the most challenging scenario for knowledge transfer due to its highest level of heterogeneity. Even so, TransNEST slightly outperforms SSVD overall and shows a clear advantage over SSVD for rare and transfer-eligible features, although such features are relatively few in this configuration.

In Table \ref{table2}, we not only focus on the angles between two embeddings but also on the scales. 
We report F.err, defined as $n_2^{-1} \|\blM_{2} - \widehat \blS_2 \|_F$, where $\widehat \blS_2  = \widehat \blX_2\widehat\blX_2\trans$; F.rare.err, defined as $n_r^{-1} \|\blM_{2, {\rm rare}, {\rm rare}} - \widehat \blS_{2, {\rm rare}, {\rm rare}} \|_F$, where $\blM_{2, {\rm rare}, {\rm rare}}$ and $\widehat \blS_{2, {\rm rare}, {\rm rare}}$ are sub-matrices of $\blM_2$ and $\widehat \blS_2$ with only the rows and columns corresponding to the rare features selected, respectively; and F.freq.err, defined as $n_f^{-1} \|\blM_{2, {\rm freq}, {\rm freq}} - \widehat \blS_{2, {\rm freq}, {\rm freq}} \|_F$, where $\blM_{2, {\rm freq}, {\rm freq}}$ and $\widehat \blS_{2, {\rm freq}, {\rm freq}}$ are sub-matrices of $\blM_2$ and $\widehat \blS_2$ with only the rows and columns corresponding to the frequent features selected, respectively.
Generally speaking, TransNEST has the best performance in $(C1)$ and $(C2)$, and comparable performance with DP in $(C3)$ and $(C4)$, outperforming the rest methods.
TransNEST significantly outperforms SSVD in the four configurations in terms of F.err and F.rare.err. It also has a smaller F.freq.err in the more homogeneous settings $(C1)$ and $(C2)$.
TransNEST outperforms SSG in terms of F.err and F.freq.err, especially in $(C3)$ and $(C4)$, where the within-group heterogeneity is larger. 
BONMI has the smallest F.freq.err but the second largest F.rare.err in the four configurations, potentially caused by the wrong assumption that all overlapping features are site-homogeneous, and the lack of utilization of prior feature-site weights.
DP has a comparable performance to TransNEST in $(C3)$ and $(C4)$, potentially due to TransNEST's failure to effectively utilize group structure to boost efficiency when the group heterogeneity is large.

The results of the remaining four configurations with $G=250$ can be found in Section J  of the Supplementary Materials, which show similar patterns. 

\begin{table*}
{\footnotesize
\caption{The performances of the embeddings from five methods on identifying the positive feature pairs against random/negative pairs.
Reported AUCs are computed using the same evaluation procedure, with the set of positive pairs restricted to all labeled available pairs (AUC), frequent labeled pairs (AUC.freq), rare labeled pairs (AUC.rare), frequent transfer-eligible labeled pairs (AUC.freq.Tr), frequent transfer-ineligible labeled pairs (AUC.freq.NTr), rare transfer-eligible labeled pairs (AUC.rare.Tr), and rare transfer-ineligible labeled pairs (AUC.rare.NTr).
}
\label{table1}
\begin{tabular}{l|lc|cccccc}
\hline
                &      & AUC & AUC.freq & AUC.rare & AUC.freq.Tr & AUC.freq.NTr& AUC.rare.Tr & AUC.rare.NTr  \\ \hline
 $(C1)$   & SSVD  &0.72 &0.78 &0.66 &0.83 &0.82 &0.66 &0.66  \\
$G = 400$      & SSG  &0.73 &0.69 &0.78 &0.87 &0.63 &0.81 &0.80    \\
$n_\zeta^c = 1000$  & DP &0.69 &0.66 &0.77 &0.88 &0.58 &0.82 &0.79  \\
$n_\delta= 100$       & BONMI &0.70 &0.69 &0.74 &0.87 &0.62 &0.76 &0.75  \\
& TransNEST   & 0.77 &0.76 &0.79 &0.87 &0.76 &0.80 &0.77\\\hline
 $(C2)$   & SSVD  &0.72 &0.75 &0.69 &0.83 &0.77 &0.66 &0.71  \\
$G = 400$      & SSG  &0.69 &0.67 &0.68 &0.87 &0.61 &0.80 &0.62    \\
$n_\zeta^c = 1000$  & DP &0.65 &0.64 &0.64 &0.88 &0.57 &0.82 &0.57  \\
$n_\delta= 200$       & BONMI &0.67 &0.67 &0.66 &0.88 &0.61 &0.77 &0.60  \\
& TransNEST &0.76 &0.75 &0.75 &0.87 &0.74 &0.79 &0.72 \\\hline
 $(C3)$   & SSVD   &0.72 &0.73 &0.67 &0.73 &0.77 &0.68 &0.68 \\
$G = 400$      & SSG & 0.70 &0.66 &0.74 &0.72 &0.62 &0.76 &0.76      \\
$n_\zeta^c = 0$  & DP &0.70 &0.67 &0.74 &0.78 &0.59 &0.77 &0.75  \\
$n_\delta= 100$       & BONMI &0.70 &0.68 &0.73 &0.77 &0.62 &0.75 &0.74 \\
& TransNEST &0.75 &0.73 &0.75 &0.74 &0.74 &0.77 &0.75\\\hline
 $(C4)$   & SSVD & 0.72 &0.73 &0.69 &0.73 &0.74 &0.67 &0.70   \\
$G = 400$      & SSG  &0.67 &0.65 &0.67 &0.71 &0.61 &0.76 &0.61    \\
$n_\zeta^c = 0$  & DP &0.66 &0.65 &0.64 &0.78 &0.58 &0.76 &0.58  \\
$n_\delta= 200$       & BONMI  &0.68 &0.67 &0.66 &0.77 &0.61 &0.75 &0.61\\
& TransNEST &0.73 &0.71 &0.71 &0.74 &0.71 &0.76 &0.69\\\hline
\end{tabular}}
\end{table*}
\begin{table*}
\centering
\caption{The performances of five methods on estimating the target error-free matrix $\blM_2$, $\blM_{2, {\rm rare}, {\rm rare}} $, and $\blM_{2, {\rm freq}, {\rm freq}} $ in terms of Frobenius errors.}
\label{table2}
\begin{tabular}{lccc|ccc}
\hline
          & F.err & F.rare.err & F.freq.err & F.err & F.rare.err & F.freq.err \\\hline
          & \multicolumn{3}{c}{$(C1)$}        & \multicolumn{3}{c}{$(C2)$}        \\\hline
SSVD      &16.11 &34.24 &8.60      &16.00 &33.86 &8.31           \\
SSG       &10.23 &13.80 &9.12            &10.24 &13.13 &9.04                  \\
DP        &11.73 &19.61 &8.28       & 11.64 &18.46 &8.54        \\
BONMI     &13.86 &28.37 &7.91       &13.53 &26.86 &7.82       \\
TransNEST &9.65 &13.06  &8.23         &9.84  &13.55 &8.23           \\\hline
          & \multicolumn{3}{c}{$(C3)$}        & \multicolumn{3}{c}{$(C4)$}        \\\hline
SSVD      &  15.53 &31.64 &8.65          &  15.48 &31.37 & 8.63        \\
SSG       & 11.72 &14.17 &10.69          & 12.07  &14.89  &10.75         \\
DP        &11.13 &17.29 &8.42             &11.58 &17.57 &8.75           \\
BONMI     &12.26  &23.14 &7.68           & 12.18 &22.30 &7.68       \\
TransNEST &11.35 &14.29 &9.91            & 11.81 &15.26 &10.14          \\\hline
\end{tabular}
\end{table*}

\section{Real Data Analysis}
The analysis included data from a total of 0.25 million patients at Boston Children's Hospital (BCH), a quaternary referral center for pediatric care, and 2.5 million patients at Mass General Brigham (MGB),  a Boston-based non-profit hospital system serving primarily an adult population, each of whom had at least one visit with codified medical records.
We gathered four domains of codified features from BCH and MGB, including PheCodes for diagnosis, RxNorm codes for medication, LOINC codes for lab measurements, and CCS codes for procedures.
After frequency control, 3055 codes are shared between the two hospital systems, while 1221 codes are unique to BCH, and 2350 codes are unique to MGB. 
We leverage the top-down hierarchical structures of PheCodes, RxNorm codes, and LOINC codes as prior group knowledge. 
More details can be found in \citep{li2024multisource}.
We then generated a summary-level co-occurrence matrix and corresponding SPPMI matrix of EHR codes at each site as described in \citep{beam2020clinical}. 

Similar to the simulation studies, we compare our TransNEST method with four benchmarks: SSVD, SSG, DP, and BONMI.
For evaluation, we curated labels from pediatric articles on reputable sources, covering both PheCode to PheCode and PheCode to RxNorm code pairs.
We randomly selected a subset of the labels for tuning and the remaining for evaluation. 
Similar to the simulation studies, we also randomly selected the same number of code pairs and treated them as negative pairs, i.e., the two codes are unrelated, based on the fact that the network is very sparse.
We assessed the quality of the embeddings by measuring the accuracy of the cosine similarity between two embeddings of each code pair, and reported AUCs for different types of codes. 
To further delineate the advantages of TransNEST, we categorized BCH EHR codes into two groups based on their occurrence frequencies within patient records: rare codes, seen in less than 0.1\% of patients, and frequent codes, observed in 0.1\% or more of patients. 

The results are summarized in Table \ref{table3}. 
Among the two single-site methods, SSG outperforms SSVD, highlighting the utility of incorporating group structures.
DP, BONMI, and TransNEST all outperform SSVD, suggesting a degree of similarity between MGB and BCH that allows effective knowledge transfer from MGB to enhance pediatric embeddings.
Furthermore, TransNEST outperforms BONMI by effectively modeling site heterogeneity in overlapping codes and leveraging group structures.
It achieves performance comparable to DP in identifying frequent PheCode–PheCode pairs but substantially higher AUCs in identifying PheCode–RxNorm pairs and rare PheCode–PheCode pairs.
This suggests that frequent PheCodes are more similar between MGB and BCH than RxNorm codes. For rare BCH codes, the combined benefits of group structure, transferred knowledge from MGB, and adaptive modeling of heterogeneity are critical, enabling TransNEST to excel.

\begin{table*}
\centering
\caption{AUCs in identifying different types of positive pairs against random/negative pairs. P-P represents PheCode-PheCode pairs, P-R represents PheCode-RxNorm pairs, P-P freq represents PheCode-PheCode pairs with at least one code being frequent, P-R freq represents PheCode-RxNorm pairs with at least one code being frequent,  P-P rare represents PheCode-PheCode pairs with at least one code being rare, and  P-R rare represents PheCode-PheCode pairs with at least one code being rare.}
\label{table3}
\begin{tabular}{l|cc|cccc}
\hline
          & P-P & P-R & P-P freq & P-R freq & P-P rare & P-R rare       \\\hline
Number of pairs &3715 &339 &3649 & 319 & 797 &138  \\ \hline          
SSVD      &0.75     &0.70     &0.74          & 0.71         &0.60          &0.63          \\
SSG       &0.77     &0.75     &0.76          &0.75          &0.71          &0.71          \\
DP        &0.81     &0.77     &0.79          & 0.77        & 0.76         & 0.74         \\
BONMI     & 0.79    & 0.72    & 0.77         &  0.74        & 0.74         &  0.68        \\
TransNEST & 0.80    & 0.81   & 0.78         & 0.80         & 0.80        & 0.80         \\ \hline
\end{tabular}
\end{table*}


\section{Discussion}

We propose a novel transfer learning under a structured missingness framework, TransNEST, based on network embeddings. It leverages prior hierarchical structures to enhance knowledge transferability across sites and adaptively captures complex heterogeneity both within groups and between sites in a robust manner.
TransNEST is also intrinsically federated, preserving data privacy by communicating only summary-level information across sites, e.g., code co-occurrence matrices of patient cohorts, without sharing patient-level data.
With the advancement of LLMs, high-quality source or general-purpose embeddings have become widely accessible. However, these embeddings may be too generic for specific downstream tasks.
TransNEST offers a principled way to fine-tune such embeddings, enabling them to better capture target-specific characteristics and statistical patterns.
More importantly, we rigorously establish theoretical guarantees for the TransNEST method, which also sheds light on the broader theoretical analysis of low-rank models, where obtaining a $\|\cdot\|_{2\to\infty}$ error bound remains a significant challenge.

Although we focus on two sites in this paper, TransNEST can be naturally extended to a multi-source setting.
To facilitate the theoretical derivations, we adopt a hard-thresholding strategy for classifying code types. An interesting direction for future work is to explore a soft-thresholding approach, which may better capture the continuously varying levels of heterogeneity within groups and between sites.
Another promising direction is to more effectively leverage prior multi-layer hierarchical structures, potentially through the use of Graph Neural Networks.




\begin{supplement}
\stitle{Supplementary Materials to `Transfer Learning with Network Embeddings under Structured Missingness'}
\sdescription{
Technical assumptions, intermediate theoretical results, proofs, and additional simulation results are provided in the Supplementary Materials.
}
\end{supplement}


\bibliographystyle{imsart-number} 
\bibliography{ref}       

\end{document}